\documentclass{article} % For LaTeX2e
\usepackage{arxiv}

\usepackage{amsmath,amsfonts,bm}

\def\eqref#1{equation~\ref{#1}}
\def\1{\bm{1}}

\DeclareMathAlphabet{\mathsfit}{\encodingdefault}{\sfdefault}{m}{sl}
\SetMathAlphabet{\mathsfit}{bold}{\encodingdefault}{\sfdefault}{bx}{n}

\usepackage{hyperref}
\usepackage{url}
\usepackage{amsmath, amsthm, amssymb, epsfig, algorithm}

\usepackage{color}

\title{Federated Quantum Natural Gradient Descent for Quantum Federated Learning}

\author{ 
Jun Qi\\
Georgia Institute of Technology\\
jqi41@gatech.edu
}
\renewcommand{\shorttitle}{Federated Quantum Natural Gradient Descent for Quantum Federated Learning}

\begin{document}

\maketitle

\begin{abstract}
The heart of Quantum Federated Learning (QFL) is associated with a distributed learning architecture across several local quantum devices and a more efficient training algorithm for the QFL is expected to minimize the communication overhead among different quantum participants. In this work, we put forth an efficient learning algorithm, namely federated quantum natural gradient descent (FQNGD), applied in a QFL framework which consists of the variational quantum circuit (VQC)-based quantum neural networks (QNN). The FQNGD algorithm admits much fewer training iterations for the QFL model to get converged and it can significantly reduce the total communication cost among local quantum devices. Compared with other federated learning algorithms, our experiments on a handwritten digit classification dataset corroborate the effectiveness of the FQNGD algorithm for the QFL in terms of a faster convergence rate on the training dataset and higher accuracy on the test one.
\end{abstract}

\section{Introduction}
\label{sec1}

Successful deep learning (DL) applications, such as automatic speech recognition (ASR)~\cite{huang2014historical}, natural language processing (NLP)~\cite{hirschberg2015advances}, and computer vision~\cite{voulodimos2018deep}, highly rely on the hardware breakthrough of the graphical processing unit (GPU). However, the advancement of large DL models, such as GPT~\cite{brown2020language} and BERT~\cite{devlin2018bert}, is faithfully attributed to significantly powerful computing capabilities that are only privileged to big companies which own numerous costly and industrial-level GPUs. With recent years witnessing a rapid development of near-term intermediate-scale quantum (NISQ) devices~\cite{Preskill2018quantumcomputingin, egan2021fault, guo2021testing}, the quantum computing hardware is expected to further speed up the classical DL algorithms by designing novel quantum machine learning (QML) models like quantum neural networks (QNN) on NISQ devices~\cite{cerezo2021variational, power_data, du2021learnability, huang2022quantum}. Unfortunately, two obstacles prevent the NISQ devices from applying to QNN in practice. The classical DL models cannot be directly deployed on a NISQ device without conversion to a quantum tensor format~\cite{qi2021qtn, chen2021end}. For another, the NISQ devices admit a few physical qubits such that insufficient qubits could be spared for the quantum error correction~\cite{ball2021real, egan2021fault, guo2021testing}, and more significantly, the representation power of QML is quite limited due to the small number of currently available qubits~\cite{qi2022theoretical}. 

To deal with the first challenge, the variational algorithm, namely the variational quantum circuit (VQC), was proposed to enable QNN simulated on NISQ devices and has attained even exponentially advantages over the DL counterparts on exclusively many tasks like ASR~\cite{yang2021decentralizing, qi2021classical}, NLP~\cite{yang2022bert}, and reinforcement learning~\cite{chen2020variational}. As for the second challenge, distributed quantum machine learning systems, which consist of different quantum machines with limited quantum capabilities, can be set up to increase the quantum computing power. One particular distributed model architecture is called quantum federated learning (QFL), which aims at a decentralized computing architecture derived from classical federated learning (FL). 

\begin{figure}[htbp]
\centerline{\includegraphics[width=135mm]{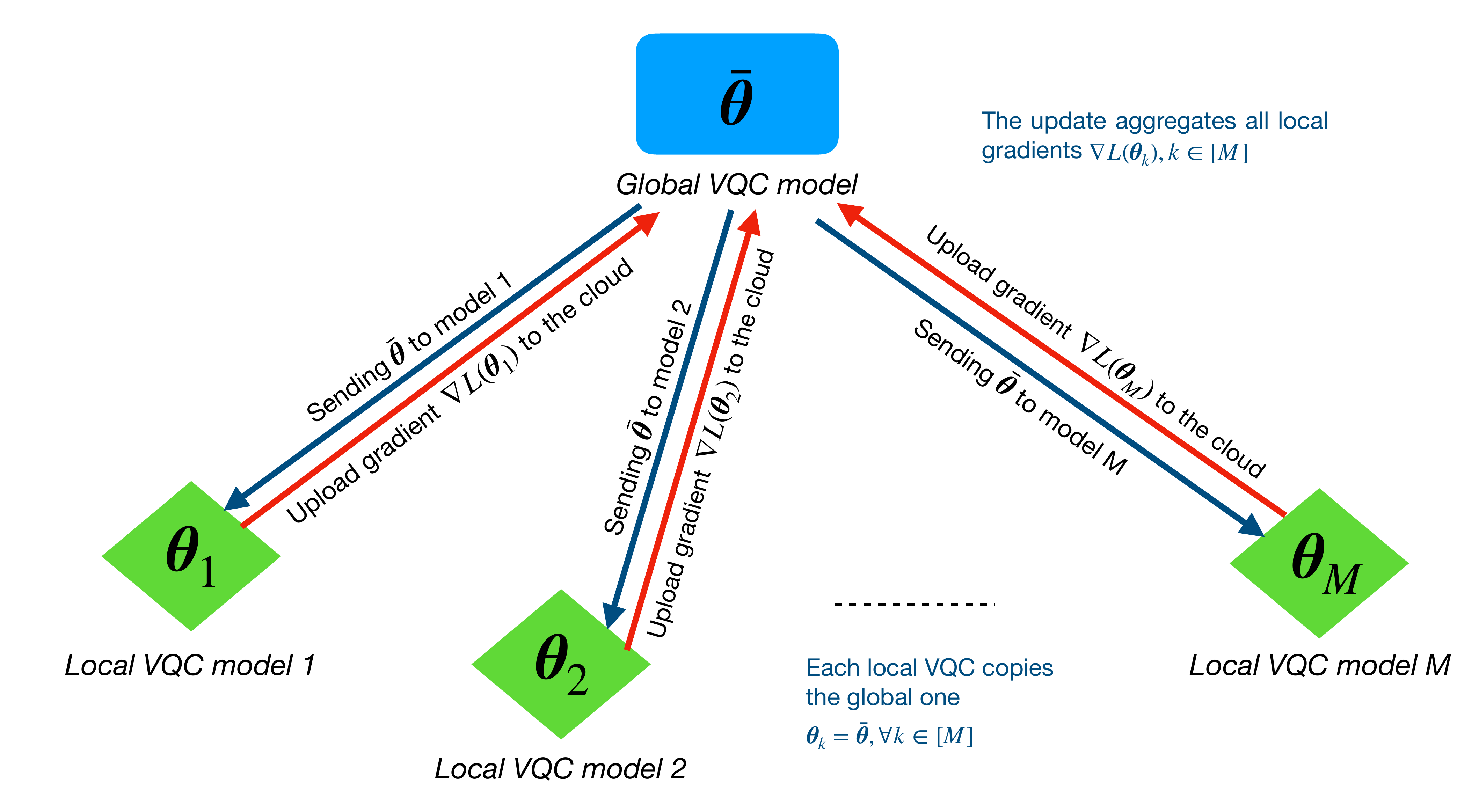}}
\caption{{\it An illustration of quantum federated learning. The global VQC parameter $\bar{\boldsymbol{\theta}}$ is first transmitted to local VQCs $\boldsymbol{\theta}_{k}$. Then, the updated gradients $\nabla\mathcal{L}(\boldsymbol{\theta}_{k})$ based on the participants' local data are sent back to the centralized server and then they are aggregated to update the parameters of the global VQC.}}
\label{fig:qfl}
\end{figure}

The FL framework depends on the advances in hardware progress, making tiny DL systems practically powerful. For example, a speech recognition system on the cloud can transmit a global acoustic model to a user's cell phone and then send the updating information back to the cloud without collecting the user's private data on the centralized computing server. This methodology helps to build a privacy-preserving DL system and inspires us to leverage quantum computing to expand machine learning capabilities. As shown in Figure~\ref{fig:qfl}, our proposed QFL and FL differ in the models utilized in the systems, where QFL is composed of VQC models rather than the classical DL counterparts for FL. More specifically, the QFL system consists of a global VQC model placed on the cloud, and $M$ local VQC models deployed on user devices. Moreover, the training process of QFL involves three key steps: (1) the global VQC parameter $\bar{\boldsymbol{\theta}}$ is transmitted to $K$ local participants' devices; (2) the parameter of each local VQC is adaptively trained based on the participant's data and then uploads the model gradients $\nabla\mathcal{L}(\boldsymbol{\theta}_{k})$ back to the cloud; (3) the uploaded gradients are aggregated to generate a centralized gradient to update the global model parameter $\bar{\boldsymbol{\theta}}$. 

Besides, a significantly inherent obstacle of QFL is bound up with the communication overhead among different VQC models. To reduce the communication cost, a more efficient training algorithm is expected to accelerate the convergence rate such that fewer counts of the global model update can be obtained. Therefore, in this work, we put forth a federated quantum learning algorithm for training QFL, namely federated quantum natural gradient descent (FQNGD). FQNGD is derived from the algorithm of quantum natural gradient descent (QNGD), which has demonstrated more efficient performance for training VQC models than the other stochastic gradient descent (SGD) methods in the training process of a single VQC.

\subsection{Main Results}
Our main contributions to this work are summarized as follows:
\begin{enumerate}
\item We design the FQNGD algorithm for the QFL system by extending the QNGD method for training a single VQC. The QNGD algorithm is developed by approximating the Fubini-Study metric tensor to create a specific VQC model structure. 

\item We compare our FQNGD algorithm with the conventional SGD algorithms, and highlight the performance advantages of FQNGD over other SGD methods in theory. 

\item Empirical study of experiments on the handwritten digit classification is conducted to corroborate our theoretical results. The experimental results can demonstrate the performance of the advantages of FQNGD for QFL. 
\end{enumerate}

\section{Related Work}

Kone{\v{c}}n{\`y} \emph{et al.}~\cite{konevcny2016federated} first proposed the FL strategies to improve the communication efficiency of a distributed computing system, and McMahan \emph{et al.}~\cite{mcmahan2017communication} set up the FL systems with the concerns in the use of big data and a large-scale cloud-based DL~\cite{shokri2015privacy}. Chen \emph{et al.}~\cite{chen2021federated} demonstrates the QFL architecture that is built based on the classical FL paradigm, where the central node holds a global VQC and receives the trained VQC parameters from participants' local quantum devices. The algorithm of QNGD was discussed to provide an efficient training method for a single VQC~\cite{stokes2020quantum}. In particular, Stokes \emph{et al.}~\cite{stokes2020quantum} firstly claimed that the Fubini-Study metric tensor can be used for the QNGD. 

This work proposes the algorithm of FQNGD by extending the QNGD to the FL setting and highlighting the learning efficiency of FQNGD in a QFL architecture. Besides, compared with the work~\cite{chen2021federated}, in this work, the gradients of VQC are uploaded to the global model rather than the VQC parameters of local devices such that the updated gradients can be collected without getting access to the VQC parameters as shown in~\cite{chen2021federated}.

\section{Results}
\subsection{Preliminaries}
In this section, we first show the necessary preliminaries that compose the algorithmic foundations for FQNGD. More specifically, we first introduce the detailed VQC framework and then explicitly explain the steps of QNGD. 

\subsubsection{Variational Quantum Circuit}

\begin{figure}
\centerline{\includegraphics[width=135mm]{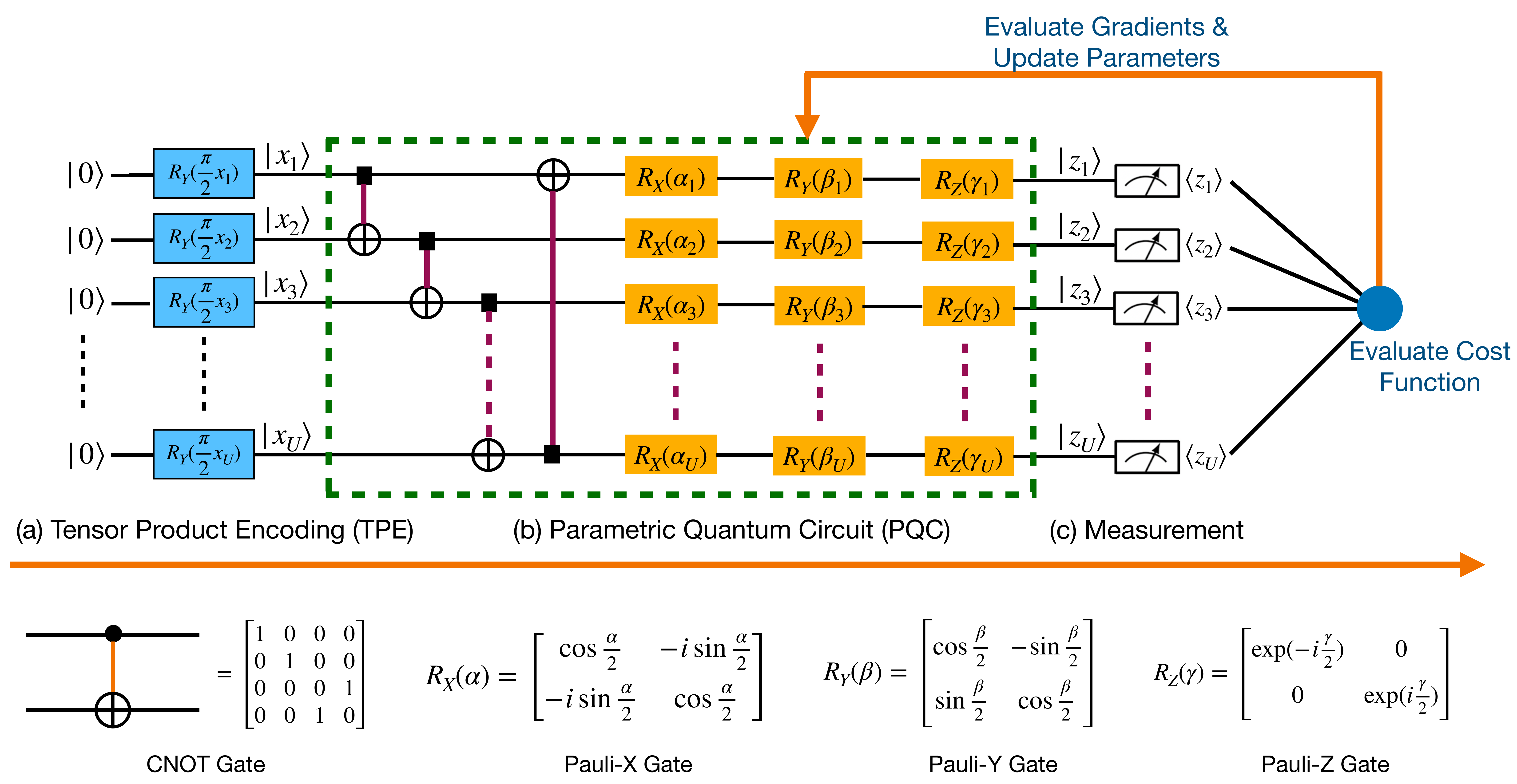}}
\caption{{\it The VQC is composed of three components: (a) TPE; (b) PQC; (c) measurement. The TPE utilizes a series of $R_{Y}(\frac{\pi}{2} x_{i})$ to transform classical inputs into quantum states. The PQC consists of CNOT gates and single-qubit rotation gates $R_{X}$, $R_{Y}$, $R_{Z}$ with trainable parameters $\boldsymbol{\alpha}$, $\boldsymbol{\beta}$, and $\boldsymbol{\gamma}$. The CNOT gates are non-parametric and impose the property of quantum entanglement among qubits, and $R_{X}$, $R_{Y}$ and $R_{Z}$ are parametric gates and can be adjustable during the training stage. To build a deep model, the PQC model in the green dash square is repeatably copied. The measurement converts the quantum states $\vert z_{1}\rangle, \vert z_{2}\rangle, ..., \vert z_{U}\rangle$ into the corresponding expectation values $\langle z_{1} \rangle, \langle z_{2} \rangle, ..., \langle z_{U} \rangle$. The outputs $\vert z_{1}\rangle$, $\vert z_{2}\rangle$, ..., $\vert z_{U}\rangle$ are connected to a loss function and the gradient descent algorithms can be used to update the VQC model parameters. Besides, both CNOT gates and $R_{X}$, $R_{Y}$ and $R_{Z}$ correspond to unitary matrices as shown below the VQC framework.}}
\label{fig:vqc}
\end{figure}

An illustration of VQC is shown in Figure~\ref{fig:vqc}, where the VQC model consists of three components: (a) tensor product encoding (TPE); (b) parametric quantum circuit (PQC); (c) measurement. The TPE initializes the input quantum states $\lvert \textbf{x}_{1} \rangle$, $\lvert \textbf{x}_{2} \rangle$, ..., $\lvert \textbf{x}_{U} \rangle$ from the classical inputs $\textbf{x}_{1}$, $\textbf{x}_{2}$, ..., $\textbf{x}_{U}$, and the PQC operator transforms the quantum states $\lvert \textbf{x}_{1} \rangle$, $\lvert \textbf{x}_{2} \rangle$, ..., $\lvert \textbf{x}_{U} \rangle$ into the output quantum states $\lvert \textbf{z}_{1} \rangle$, $\lvert \textbf{z}_{2} \rangle$, ..., $\lvert \textbf{z}_{U} \rangle$. The measurement outputs the expected observations $\langle \textbf{z}_{1} \rangle$, $\langle\textbf{z}_{2} \rangle$, ..., $\langle \textbf{z}_{U} \rangle$. 

In more detail, the TPE model was firstly proposed in \cite{stoudenmire2016supervised} and aims at converting a classical vector $\textbf{x}$ into a quantum state $\lvert \textbf{x} \rangle$ by building up a one-to-one mapping as:
\begin{equation}
\label{eq:tpe}
\vert \textbf{x} \rangle = \left(\otimes_{i=1}^{U} R_{Y}(\frac{\pi}{2} x_{i}) \right) \vert 0 \rangle^{\otimes U} = \begin{bmatrix} \cos(\frac{\pi}{2} x_{1}) \\ \sin(\frac{\pi}{2} x_{1}) \end{bmatrix} \otimes \begin{bmatrix} \cos(\frac{\pi}{2} x_{2}) \\ \sin(\frac{\pi}{2} x_{2}) \end{bmatrix} \otimes \cdot\cdot\cdot \otimes \begin{bmatrix} \cos(\frac{\pi}{2} x_{U}) \\ \sin(\frac{\pi}{2} x_{U}) \end{bmatrix},
\end{equation}
where $R_{Y}(\cdot)$ refers to a single-qubit quantum gate rotated across $Y$-axis and each $x_{i}$ is constraint to the domain of $[0, 1]$ that results in a reversely one-to-one conversion between $\textbf{x}$ and $\vert \textbf{x} \rangle$. 

Moreover, the PQC is equipped with the CNOT gates for quantum entanglement and trainable quantum gates $R_{X}(\alpha_{i})$, $R_{Y}(\beta_{i})$, and $R_{Z}(\gamma_{i})$, where the qubit angles $\alpha_{i}$, $\beta_{i}$, and $\gamma_{i}$ are adjustable during the training process. The PQC framework in the green dash square is repeatedly copied to set up a deep model, and the number of the copied PQC frameworks is called the depth of VQC. The operation of the measurement outputs the classical expected observations $\vert z_{1}\rangle$, $\vert z_{2}\rangle$, ..., $\vert z_{U}\rangle$ from the quantum output states. The expected outputs are used to calculate the loss value and the gradient descents~\cite{ruder2016overview} that can be utilized to update the VQC model parameters by applying the back-propagation algorithm~\cite{werbos1990backpropagation}.

\subsubsection{Quantum Natural Gradient Descent}

As shown in Eq. (\ref{eq:gd}), at the step $t$, the standard gradient descent minimizes a loss function $\mathcal{L}(\boldsymbol{\theta})$ with respect to the parameters $\boldsymbol{\theta}$ in a Euclidean space. 
\begin{equation}
\label{eq:gd}
\boldsymbol{\theta}_{t+1} = \boldsymbol{\theta}_{t} - \eta \nabla \mathcal{L}(\boldsymbol{\theta}_{t}). 
\end{equation}

The standard gradient descent algorithm admits each optimization step conducted in a Euclidean geometry on the parameter space. However, since the form of parameterization is not unique, different compositions of parameterizations are likely to distort distances within the optimization landscape. A better alternative method is to perform the gradient descent in the distribution space, namely natural gradient descent~\cite{amari1998natural}, which is dimension-free and invariant concerning parameterization. In doing so, each optimization step chooses the optimum step size for the update of parameter $\boldsymbol{\theta}_{t}$, regardless of the choice of parameterization. Mathematically, the standard gradient descent is modified as follows:
\begin{equation}
\label{eq:fisher}
\boldsymbol{\theta}_{t+1} = \boldsymbol{\theta}_{t} - \eta F^{-1}\nabla \mathcal{L}(\boldsymbol{\theta}_{t}), 
\end{equation}
where $F$ denotes the Fisher information matrix that acts as a metric tensor, transforming the steepest gradient descent in the Euclidean parameter space to the steepest descent in the distribution space. 

Since the standard Euclidean geometry is sub-optimal for the optimization of quantum variational algorithms, a quantum analog has the following form:
\begin{equation}
\label{eq:nw}
\boldsymbol{\theta}_{t+1} = \boldsymbol{\theta}_{t} - \eta g^{+}(\boldsymbol{\theta}_{t})\nabla\mathcal{L}(\boldsymbol{\theta}_{t}), 
\end{equation}
where $g^{+}(\boldsymbol{\theta}_{t})$ refers to the pseudo-inverse and is associated with the specific architecture of the quantum circuit. The coefficient $g^{+}(\boldsymbol{\theta}_{t})$ can be calculated using the Fubini-Study metric tensor and it reduces to the Fisher information matrix in the classical limit~\cite{mcardle2019variational}.

\subsection{Theoretical Results}

\subsubsection{Quantum Natural Gradient Descent for the VQC}

Before employing the QFNGD for a quantum federated learning system, we concentrate on the use of QNGD for a single VQC. For simplicity, we leverage a block-diagonal approximation to the Fubini-Study metric tensor for composing QNGD for the training of the VQC on the NISQ quantum hardware. 

Given an initial quantum state $\lvert \psi_{0} \rangle$ and a PQC with $L$ layers, for $l\in [L], $we separately denote $\textbf{W}_{l}$ and $\textbf{V}_{l}(\boldsymbol{\theta}_{l})$ as the unitary matrices associated with non-parameterized quantum gates and parameterized quantum ones. 

\begin{figure}[htbp]
\centerline{\includegraphics[width=135mm]{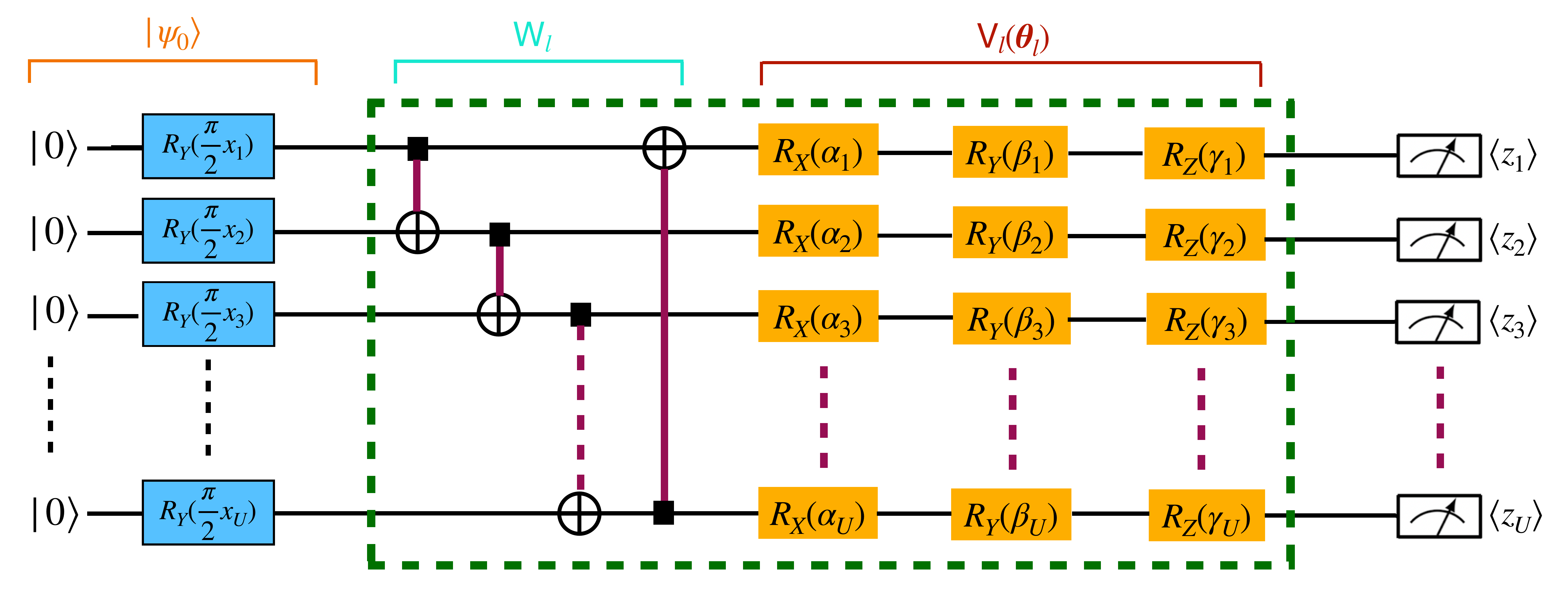}}
\caption{{\it An illustration of unitary matrices associated with the non-parametric and parametric gates. $\forall l\in [L]$, the matrices $\textbf{W}_{l}$ correspond to the non-parametric gates, the matrices $\textbf{V}_{l}(\boldsymbol{\theta}_{l})$ are associated with the parametric ones, and $\lvert \psi_{0} \rangle$ refers to the initial quantum state that is derived from the operation of TPE.}}
\label{fig:fig2}
\end{figure}

Consider a variational quantum circuit as: 
\begin{equation}
U(\boldsymbol{\theta}) \lvert \psi_{0} \rangle = \textbf{V}_{L}(\boldsymbol{\theta}_{L}) \textbf{W}_{L} \cdot\cdot\cdot \textbf{V}_{l}(\boldsymbol{\theta}_{l}) \textbf{W}_{l} \cdot\cdot\cdot \textbf{V}_{1}(\boldsymbol{\theta}_{1}) \textbf{W}_{1} \lvert \psi_{0} \rangle 
\end{equation}

Furthermore, any unitary quantum parametric gates can be rewritten as $V_{l}(\boldsymbol{\theta}_{l}) = \exp(i \boldsymbol{\theta}_{l} H_{l})$, where $H_{l}$ refers to the Hermitian generator of the gate $V_{L}$. The approximation to the Fubini-Study metric tensor admits that for each parametric layer $l$ in the variational quantum circuit, the $n_{l} \times n_{l}$ block-diagonal submatrix of the Fubini-Study metric tensor $g_{l,i,j}^{+}$ is calculated by 

\begin{equation}
\label{eq:app}
g_{l,i,j}^{+} = \langle \psi_{l} \lvert H_{l}(i) H_{l}(j) \lvert \psi_{l} \rangle - \langle \psi_{l} \lvert H_{l}(i) \lvert \psi_{l} \rangle \langle \psi_{l} \vert H_{l}(j) \vert \psi_{l} \rangle,
\end{equation}
where 
\begin{equation}
\label{eq:part}
\vert \psi_{l} \rangle = \textbf{V}_{l}(\boldsymbol{\theta}_{l})  \textbf{W}_{l} \cdot\cdot\cdot \textbf{V}_{1}(\boldsymbol{\theta}_{1})\textbf{W}_{1} \vert \psi_{0} \rangle. 
\end{equation}

\begin{figure}[htbp]
\centerline{\includegraphics[width=120mm]{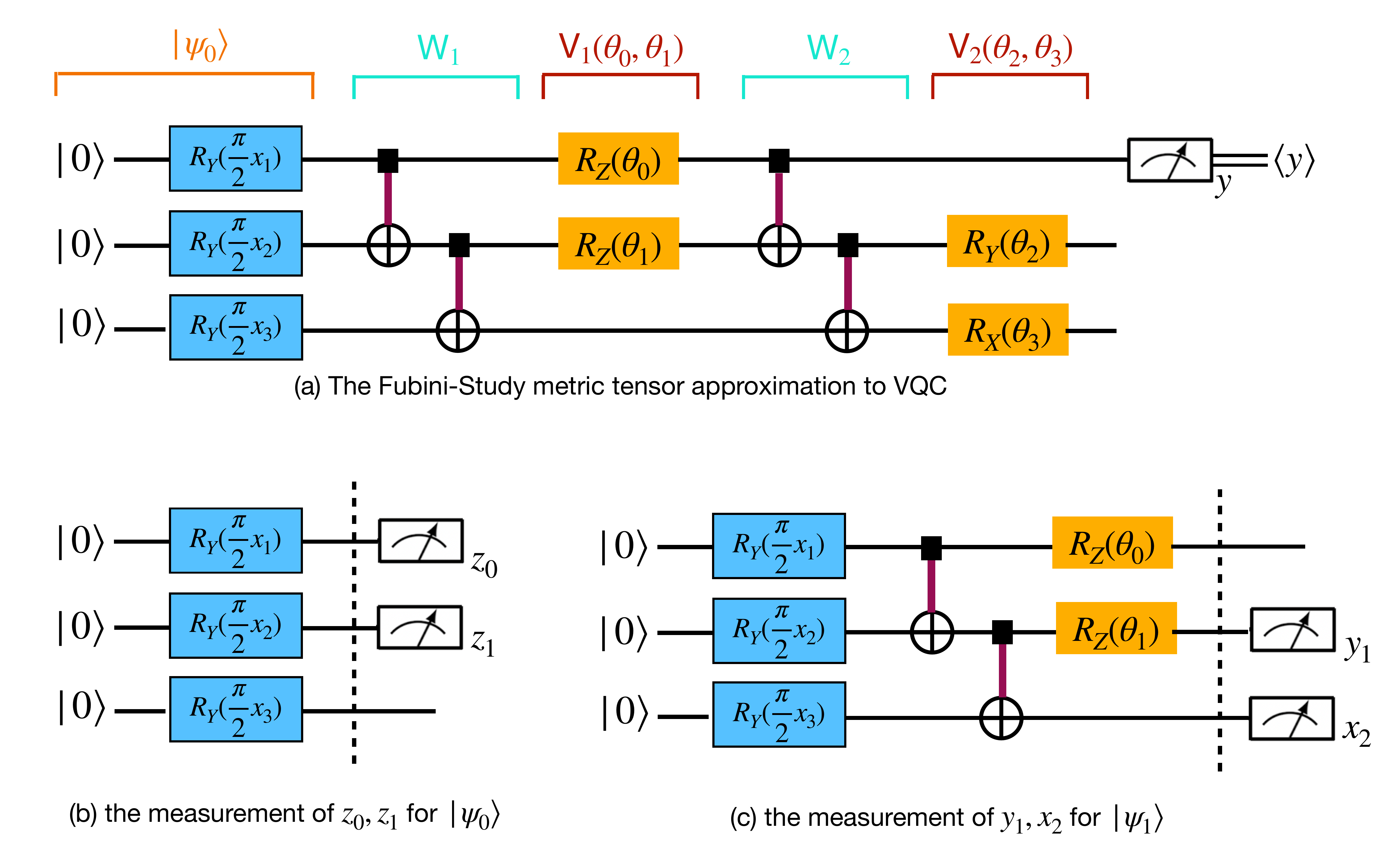}}
\caption{{\it A demonstration of the VQC approximation method based on the Fubini-Study metric tensor: (a) A block-diagonal approximation to VQC based on the Fubini-Study metric tensor; (b) a measurement of $z_{0}, z_{1}$ for $\vert \psi_{0} \rangle$; a measurement of $y_{1}, x_{2}$ for $\vert \psi_{1} \rangle$. }}
\label{fig:fig3}
\end{figure}

$\vert \psi_{l}\rangle$ denotes the quantum state before the application of the parameterized layer $l$. Figure~\ref{fig:fig3} illustrates a simplified version of VQC, where $\textbf{W}_{1}$, $\textbf{W}_{2}$ are related to non-parametric gates, and $\textbf{V}_{1}(\theta_{0}, \theta_{1})$ and $\textbf{V}_{2}(\theta_{2}, \theta_{3})$ correspond to the parametric gates with adjustable parameters , respectively. Since there are two layers, each of which owns two free parameters, the block-diagonal approximation is composed of two $2 \times 2$ matrices, $g^{+}_{1}$ and $g^{+}_{2}$. More specifically, $g^{+}_{1}$ and $g^{+}_{2}$ can be separately expressed as:

\begin{equation}
g^{+}_{1} = \begin{bmatrix}
\langle z_{0}^{2} \rangle - \langle z_{0} \rangle^{2} && \langle z_{0}z_{1} \rangle - \langle z_{0} \rangle \langle z_{1} \rangle  \\
 \langle z_{0}z_{1} \rangle - \langle z_{0} \rangle \langle z_{1} \rangle && \langle z_{1}^{2} \rangle - \langle z_{1} \rangle^{2}
\end{bmatrix},
\end{equation}
and 
\begin{equation}
g^{+}_{2} = \begin{bmatrix}
\langle y_{1}^{2} \rangle - \langle y_{1} \rangle^{2} && \langle y_{1}x_{2} \rangle - \langle y_{1} \rangle \langle x_{2} \rangle  \\
 \langle y_{1}x_{2} \rangle - \langle y_{1} \rangle \langle x_{2} \rangle && \langle x_{2}^{2} \rangle - \langle x_{2} \rangle^{2}
\end{bmatrix}.
\end{equation}

The elements of $g_{1}^{+}$ and $g_{2}^{+}$ compose $g^{+}(\boldsymbol{\theta})$ as:
\begin{equation}
g^{+}(\boldsymbol{\theta}) = \begin{bmatrix}
g_{1}^{+} & 0 \\
0 & g_{2}^{+}
\end{bmatrix}. 
\end{equation}

Then, we employ the Eq.~(\ref{eq:nw}) to update the VQC parameter $\boldsymbol{\theta}$. 

\subsubsection{Federated Quantum Natural Gradient Descent}

A QFL system can be simply built by setting up VQC models in an FL manner, where the QNGD algorithm is applied for each VQC and the uploaded gradients of all VQCs are aggregated to update the model parameters of the global VQC. In mathematical, the FQNGD can be summarized as:

\begin{equation}
\label{eq:aupdate}
\bar{\boldsymbol{\theta}}_{t+1} = \bar{\boldsymbol{\theta}}_{t} - \eta \sum\limits_{k=1}^{K} \frac{N_{k}}{N} g_{k}^{+}(\boldsymbol{\theta}^{(k)}_{t}) \nabla\mathcal{L}(\boldsymbol{\theta}^{(k)}_{t}),  
\end{equation}
where $\bar{\boldsymbol{\theta}}_{t}$ and $\boldsymbol{\theta}_{t}^{(k)}$ separately correspond to the model parameters of the global VQC and the $k$-th VQC model at epoch $t$, and $N_{k}$ represents the amount of training data stored in the participant $k$, and the sum of $K$ participants' data is equivalent to $N$.

Compared with the SGD counterparts used for QFL, the FQNGD algorithm admits adaptive learning rates for the gradients such that the convergence rate could be accelerated according to the VQC model status. 

\begin{algorithm}[tb]
   \caption{The FQNGD Algorithm for QFL}
   \label{alg:iterative}
 1. The coordinator executes. \\
 2. Initialize global parameter $\bar{\boldsymbol{\theta}}$ and broadcast it to all participants $\boldsymbol{\theta}_{0}^{(k)}$.  \\
 3. Assign each participant with $N_{k}$ training data, where $\sum_{k=1}^{K} N_{k} = N$. \\
 4. For each global model update at epoch $t=[T]$ do \\
 5. \hspace{4mm} For each participant $k \in [K]$ \textbf{in parallel} do \\
 6. \hspace{9mm} Attain $g^{+}(\boldsymbol{\theta}^{(k)}_{t}) \nabla \mathcal{L}(\boldsymbol{\theta}_{t}^{(k)})$ by applying the QNGD for the $k$-th VQC.  \\
 7. \hspace{9mm} Send the local gradient $g^{+}(\boldsymbol{\theta}^{(k)}_{t}) \nabla \mathcal{L}(\boldsymbol{\theta}_{t}^{(k)})$ to the coordinator. \\
 8. \hspace{4mm} End for \\
 9. \hspace{3.7mm} The coordinator aggregates the received gradients $g^{+}(\boldsymbol{\theta}^{(k)}_{t}) \nabla \mathcal{L}(\boldsymbol{\theta}_{t}^{(k)})$.  \\
 10.\hspace{3,4mm} The coordinator updates the global model by taking Eq. (\ref{eq:aupdate}). \\
 11. \hspace{3,3mm} Broadcast the updated global model parameter $\bar{\boldsymbol{\theta}}_{t+1}$ to all participants. \\
 12. End for
\end{algorithm}

\subsubsection{Empirical Results}

To demonstrate the FQNGD algorithm for QFL, we perform the binary and ternary classification tasks on the standard MNIST dataset~\cite{deng2012mnist}, specifically digits $\{2, 5\}$ for the binary task and $\{1, 3, 7\}$ for the ternary one. There are a total of $11379$ training data and $1924$ test data for the binary classification, and $19138$ training data and $3173$ test data are assigned for the ternary classification. As for the setup of QFL in our experiments, the QFL system consists of $6$ identically local VQC participants, each of which owns the same amount of training data. The test data are stored in the global part and are used to evaluate the classification performance. 

\begin{figure}[htbp]
\centerline{\includegraphics[width=135mm]{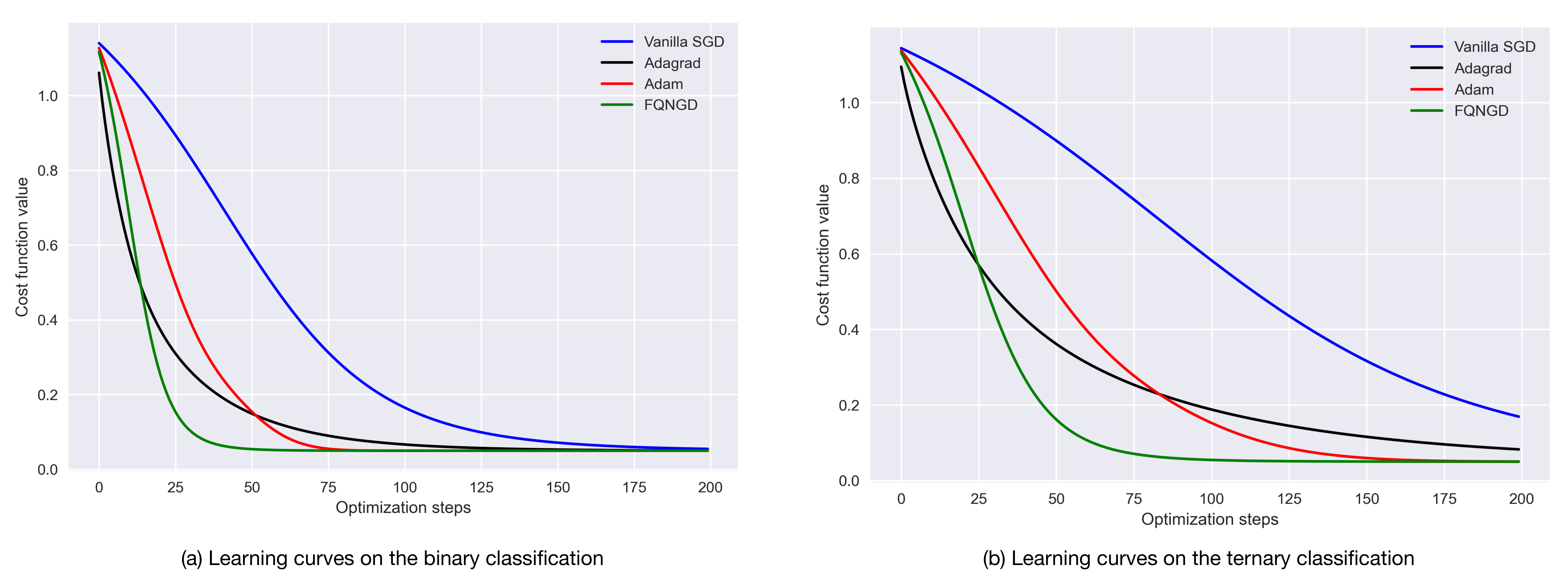}}
\caption{{\it Simulation results of binary and ternary classification on the training set of the MNIST database. (a) the learning curves of various optimization methods for the binary classification; (b) the learning curves of various optimization methods for the ternary classification.}}
\label{fig:res}
\end{figure}

The experiments compare our proposed FQNGD algorithm with the other three optimizers: the naive SGD optimizer, the Adagrad optimizer~\cite{lydia2019adagrad} and the Adam optimizer~\cite{kingma2014adam}. The Adagrad optimizer is a gradient descent optimizer with a past-gradient-dependent learning rate in each dimension. The Adam optimizer refers to the gradient descent method with an adaptive learning rate as well as adaptive first and second moments. 

\begin{table}[tpbh]\footnotesize
\center
\renewcommand{\arraystretch}{1.3}
\caption{The simulation results of a binary classification in terms of accuracy.}
\begin{tabular}{|c||c|c|c|c|}
\hline
Methods     		& SGD		&   Adagrad     	& Adam	 & FQNGD   \\
\hline
Acc.			&	98.48		&	98.81	&	98.87	&	99.32	\\
\hline
\end{tabular}
\label{tab:tab1}
\end{table}

\begin{table}[tpbh]\footnotesize
\center
\renewcommand{\arraystretch}{1.3}
\caption{The simulation results of a ternary classification in terms of accuracy.}
\begin{tabular}{|c||c|c|c|c|}
\hline
Methods     		& SGD		&   Adagrad     	& Adam	& FQNGD   \\
\hline
Acc.			& 97.86		& 98.63	&98.71	&	99.12	\\
\hline
\end{tabular}
\label{tab:tab2}
\end{table}

As shown in Figure~\ref{fig:res}, our simulation results suggest that our proposed FQNGD method is capable of achieving the fastest convergence rate compared with other optimization approaches. It means that the FQNGD method can lower the communication cost and also maintain the baseline performance of both binary and ternary classification on the MNIST dataset. Moreover, we evaluate the QFL performance in terms of classification accuracy. The FQNGD method outperforms the other counterparts with the highest accuracy values. In particular, the FQNGD is designed for the VQC model and can attain better empirical results than the Adam and Adagrad methods with adaptive learning rates over epochs.

\section{Conclusion and Discussion}
This work focuses on the design of the FQNGD algorithm for the QFL system in which multiple local VQC models are applied. The FQNGD is derived from training a single VQC based on QNGD, which relies on the block-diagonal approximation of the Fubini-Study metric tensor to the VQC architecture. We put forth the FQNGD method to train the QFL system. Compared with other SGD optimization approaches with Adagrad and Adam optimizers, our experiments on the classification tasks on the MNIST dataset demonstrate the FQNGD method attains better empirical results than other SGD counterparts, while the FQNGD exhibits a faster convergence rate than the others, which implies that our FQNGD method suggests that it is capable of lowering the communication cost and can maintain the baseline empirical results. 

Although this work focuses on the optimization methods for the QFL system, the decentralized deployment of high-performance QFL systems for adapting to the large-scale dataset is left for our future investigation. In particular, it is essential to consider how to defend against malicious attacks from adversaries and also boost the robustness and integrity of the shared information among local participants. Besides, another quantum neural network like quantum convolutional neural networks (QCNN)~\cite{cong2019quantum} is worth further study to constitute a QFL system.

\section{Method}

In our study, the Fubini-Study metric tensor is typically defined in quantum mechanics' notations of mixed states. To explicitly equate this notation to the homogeneous coordinates, let 
\begin{equation}
\vert \psi \rangle = \sum\limits_{k=0}^{K} Z_{k} \vert e_{k} \rangle = [Z_{0} : Z_{1} : ... : Z_{K}], 
\end{equation}
where $\{\vert e_{k} \rangle\}$ is a set of orthonormal basis vectors for Hilbert space, the $Z_{k}$ are complex numbers. $Z_{\alpha} = [Z_{0} : Z_{1} : ... : Z_{K}]$ is the standard notation for a point in the projective space of homogeneous coordinates. Then, given two points $\vert \psi \rangle = Z_{\alpha}$ and $\vert \phi \rangle = W_{\alpha}$ in the space, the distance between $\vert \psi \rangle$ and $\vert \phi \rangle$ is 
\begin{equation}
\gamma = \arccos \sqrt{\frac{\langle \psi \vert \phi \rangle \langle \phi \vert \psi \rangle}{ \langle \psi \vert \psi \rangle \langle \phi \vert \phi \rangle}}. 
\end{equation}

%The infinitesimal form of the Fubini-Study metric can be obtained by taking $\phi = \psi + \delta \psi$ to obtain 

%\begin{equation}
%ds^{2} = \frac{\langle \delta \psi \vert \delta \psi \rangle}{ \langle \psi \vert \psi \rangle} - \frac{\langle \delta \psi \vert \psi \rangle \langle \psi \vert \delta \psi \rangle}{\langle \psi \vert  \psi \rangle^{2}}. 
%\end{equation}

The Fubini-Study metric is the natural metric for the geometrization of quantum mechanics, and much of the peculiar behavior of quantum mechanics including quantum entanglement can be attributed to the peculiarities of the Fubini-Study metric.

\bibliography{iclr2022_conference}
\bibliographystyle{unsrt} %nat}

\end{document}